\documentclass[sigconf]{acmart}

\usepackage{booktabs} 
\usepackage{xcolor}
\usepackage{balance}

\usepackage[utf8]{inputenc}
\usepackage[TS1,T1]{fontenc}
\usepackage{booktabs} 
\usepackage{algorithm}
\usepackage{algorithmic}
\usepackage{amsmath}

\newcommand{\ds}{DataSynthesizer}
\newcommand{\dsspace}{\ds~}

\newcommand{\dd}{DataDescriber}
\newcommand{\dg}{DataGenerator}

\newcommand{\ddspace}{\dd~}
\newcommand{\dgspace}{\dg~}

\usepackage{color}

\begin{document}

\title{Synthetic Data for Social Good}

\author{Bill Howe}
\authornote{This work was supported by the University of Washington Information School, Microsoft, the Gordon and Betty Moore Foundation (Award \#2013-10-29) and the Alfred P. Sloan Foundation (Award \#3835) through the Data Science Environments program.}
\orcid{1234-5678-9012}
\affiliation{%
  \institution{University of Washington\\ Seattle, WA}
}
\email{billhowe@cs.washington.edu}

\author{Julia Stoyanovich}
\authornote{This work was supported in part by NSF Grants No. 1741047, 1464327 and 1539856, and BSF Grant No. 2014391.}
\affiliation{%
  \institution{Drexel University\\ Philadelphia, PA}
}
\email{stoyanovich@drexel.edu}

\author{Haoyue Ping}
\affiliation{%
  \institution{Drexel University\\ Philadelphia, PA}
}
\email{hp354@drexel.edu}

\author{Bernease Herman}
\affiliation{%
  \institution{University of Washington\\ Seattle, WA}
}
\email{bernease@uw.edu}

\author{Matt Gee}
\affiliation{%
  \institution{Impact Lab\\ Chicago, IL}
}
\email{mattgee@gmail.com}

\begin{abstract}

Data for good implies unfettered access to data.  But data owners must be conservative about how, when, and why they share data or risk violating the trust of the people they aim to help, losing their funding, or breaking the law.  Data sharing agreements can help prevent privacy violations, but require a level of specificity that is premature during preliminary discussions, and can take over a year to establish.
We consider the generation and use of synthetic data to facilitate ad hoc collaborations involving sensitive data.  A good synthetic dataset has two properties: it is representative of the original data, and it provides strong guarantees about privacy.  

In this paper, we discuss important use cases for synthetic data that challenge the state of the art in privacy-preserving  data generation, and describe \ds, a dataset generation tool that takes a sensitive dataset as input and generates a structurally and statistically similar synthetic dataset, with strong privacy guarantees, as output. The data owners need not release their data, while potential collaborators can begin developing models and methods with some confidence that their results will work similarly on the real dataset.  The distinguishing feature of \dsspace is its usability --- in most cases, the data owner need not specify any parameters to start generating and sharing data safely and effectively.



The code implementing \dsspace is publicly available on GitHub
at \url{https://github.com/DataResponsibly}.  The work on \dsspace is part of the Data, Responsibly project, where the goal is to operationalize responsibility in data sharing, integration, analysis and use.

\end{abstract}



\maketitle

\section{Introduction}
\label{sec:intro}


Collaborative projects in the social and health sciences increasingly require sharing sensitive, privacy-encumbered data.  Social scientists, government agencies, health workers, and non-profits are eager to collaborate with data scientists, but many projects fail before they begin due to delays incurred by data sharing agreements --- our colleagues report that 18 months is a typical timeframe to establish such agreements!   

Data scientists typically require access to data before they can understand the problem or even determine whether they can help.  But data owners cannot share data without significant legal protections in place. Beyond legal concerns, there is a general reluctance to share sensitive data with non-experts before they have ``proven themselves,'' since they typically are not familiar with the context in which the data was collected and may be distracted by spurious results.    

To bootstrap these collaborations prior to data sharing agreements being established, we advocate generating datasets that are \emph{structurally and statistically similar} to the real data but that are 1) obviously synthetic to put the data owners at ease, and 2) offer strong privacy guarantees to prevent adversaries from extracting any sensitive information.  These two requirements are not redundant: strong privacy guarantees are not always sufficient to convince data owners to release data, and even seemingly random datasets may not prevent subtle privacy attacks. With this approach, data scientists can begin to develop models and methods with synthetic data, but maintain some degree of confidence that their work will remain relevant when applied to the real data once proper data sharing agreements are in place.

As an initial exploration of these ideas, we have developed a tool named \dsspace that generates structurally and statistically similar synthetic datasets based on real, private datasets. 
Given a sensitive dataset, \dsspace infers the domain of each attribute and derives a probabilistic model of the attribute values, possibly adding noise to ensure differential privacy. The description of this derived model is saved in a \emph{dataset description file}. The tool then generates synthetic datasets of arbitrary size by sampling from the stored distribution.

\dsspace can operate in one of three modes, which differ in how the probabilistic model is derived: In \emph{correlated attribute mode}, we learn a differentially private Bayesian network capturing the correlation structure between attributes, then draw samples from this model to construct the result dataset.  In cases where the correlated attribute mode is too computationally expensive or when there is insufficient data to derive a reasonable model, one can use \emph{independent attribute mode}.  In this mode, a histogram is derived for each attribute, noise is added to the histogram to achieve differential privacy, and then samples are drawn for each attribute. Finally, for cases of extremely sensitive data, one can use \emph{random} mode that simply generates type-consistent random values for each attribute.  We give a brief overview of the implementation of the tool, and of the kinds of user interaction it supports, in Section~\ref{sec:specto}.

We envision various extensions to this basic approach. For example, joining multiple sensitive datasets requires care: the join of two synthetic datasets does not necessarily have the same properties as a synthetic dataset derived from the join of two rel datasets.  But in many cases, joins between the real data are expressly forbidden: linking education and housing datasets is important to understand the effects of homelessness on graduation rates, but FERPA laws preclude this kind of linking.  Beyond linking, we see value in mixing real data and fake data in order to adjust statistical properties or ensure anonymity requirements, adversarially generating fake datasets to assess bias and accuracy of external models, and even generating complete``cities'' of fake data based on the real data exhaust from city operations.  In the latter case, we see the resulting interconnected datasets as a research instrument that can attract researchers who may otherwise be turned off by the administrative hurdles in getting access to data.  We describe all of these extensions in more detail in Section \ref{sec:vision}.

We briefly survey related work in Section~\ref{sec:related} and conclude in Section~\ref{sec:conc}.


\section{\ds: Safe Tabular Data}
\label{sec:specto}

We instantiate our vision in an open-source tool called \ds, which takes a private dataset as input and generates synthetic datasets that can be shared safely. \dsspace generates fake data using state-of-the-art differential privacy mechanisms~\cite{DBLP:conf/tcc/DworkMNS06}.  Differential privacy is a family of techniques that guarantee that the output of an algorithm is statistically indistinguishable on a pair of \emph{neighboring} databases --- a pair of databases that differ by only one row.  That is, the presence or absence of a single individual in the input to the algorithm will be undetectable when one looks at the output.

An important property of differential privacy is that its effectiveness degrades with repeated queries~\cite{DBLP:conf/uss/HaeberlenPN11}.  To prevent leaking private information through adversaries repeatedly sending data generation requests, the system administrator can assign a unique random seed for each person who requires a synthetic dataset.  




\subsection{Overview of the implementation} 
\label{sec:unconditioned_distributions}

We now briefly describe the implementation of \ds; see Ping et al.~\cite{DBLP:conf/ssdbm/PingSH17} for additional details.  

\dsspace is implemented in Python 3. The data owner can interact with the tool through Jupyter notebooks or through a Web-based UI. \dsspace assumes that the private dataset is presented in CSV format. The tool is designed to work with minimal input from the user.  For example, it is not necessary to specify data types of the attributes, the tool determines these automatically.

The input dataset is first processed by the \ddspace module, which infers attribute data types and estimates their distributions.  For each attribute identified as categorical (one with a small number of distinct values, such as gender, ethnicity and education level), \ddspace computes the frequency distribution of each distinct value.  For non-categorical numerical and datetime attributes, \ddspace derives an equi-width histogram to represent the distribution. For non-categorical string attributes, their minimum and maximum lengths are recorded. This information is recorded in a dataset description, which is then used to generate fake datasets.  Additional processing steps may be required depending on the specific mode of operation chosen by the data owner. There are three such modes, which we describe next.


\subsubsection*{Random mode} \label{sec:build_random} 
When invoked in random mode, \dgspace generates  type-consistent random values for each attribute.   If an attribute is of type string, then a random string is generated with length that falls within the observed range of lengths in the dataset.

\subsubsection*{Independent attribute mode} \label{sec:unconditioned_distributions}

\begin{figure*}
\centering
\includegraphics[width=5.3in]{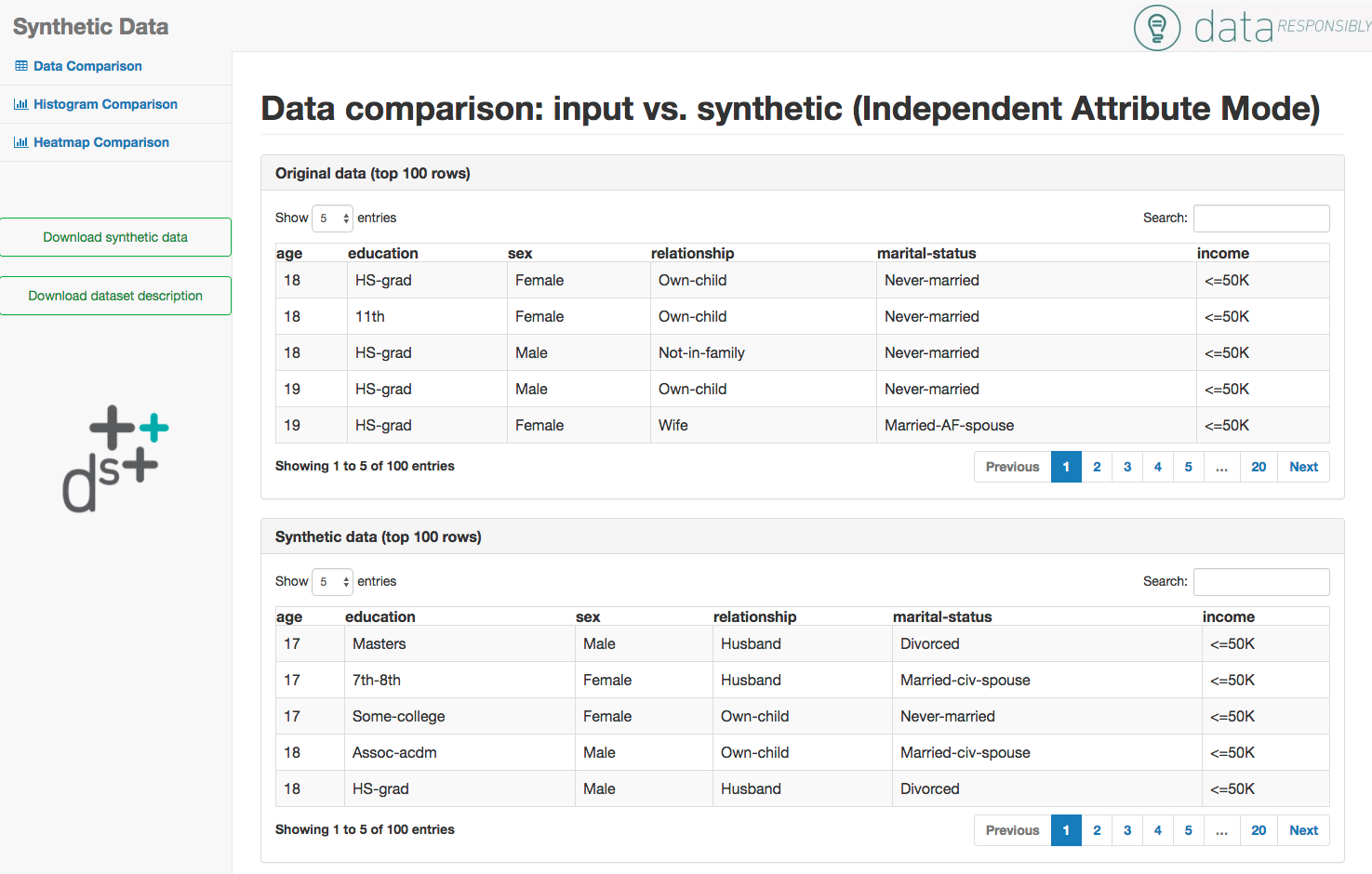}
\caption{Data comparison}
\label{fig:datacomp}
\end{figure*}

\begin{figure*}
\centering
\includegraphics[width=5.3in]{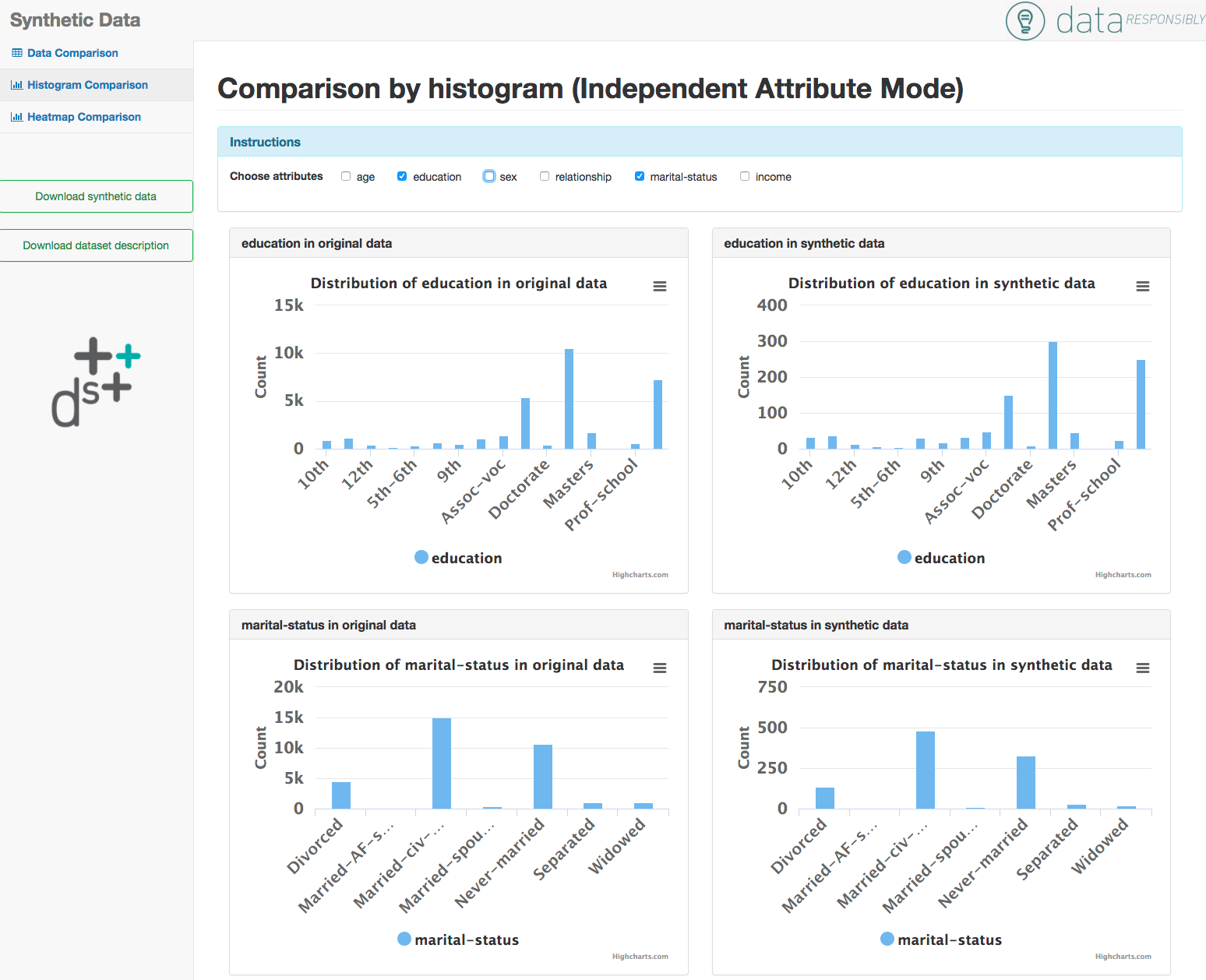}
\caption{Histogram comparison}
\label{fig:histcomp}
\end{figure*}

\begin{figure*}
\centering
\includegraphics[width=5in]{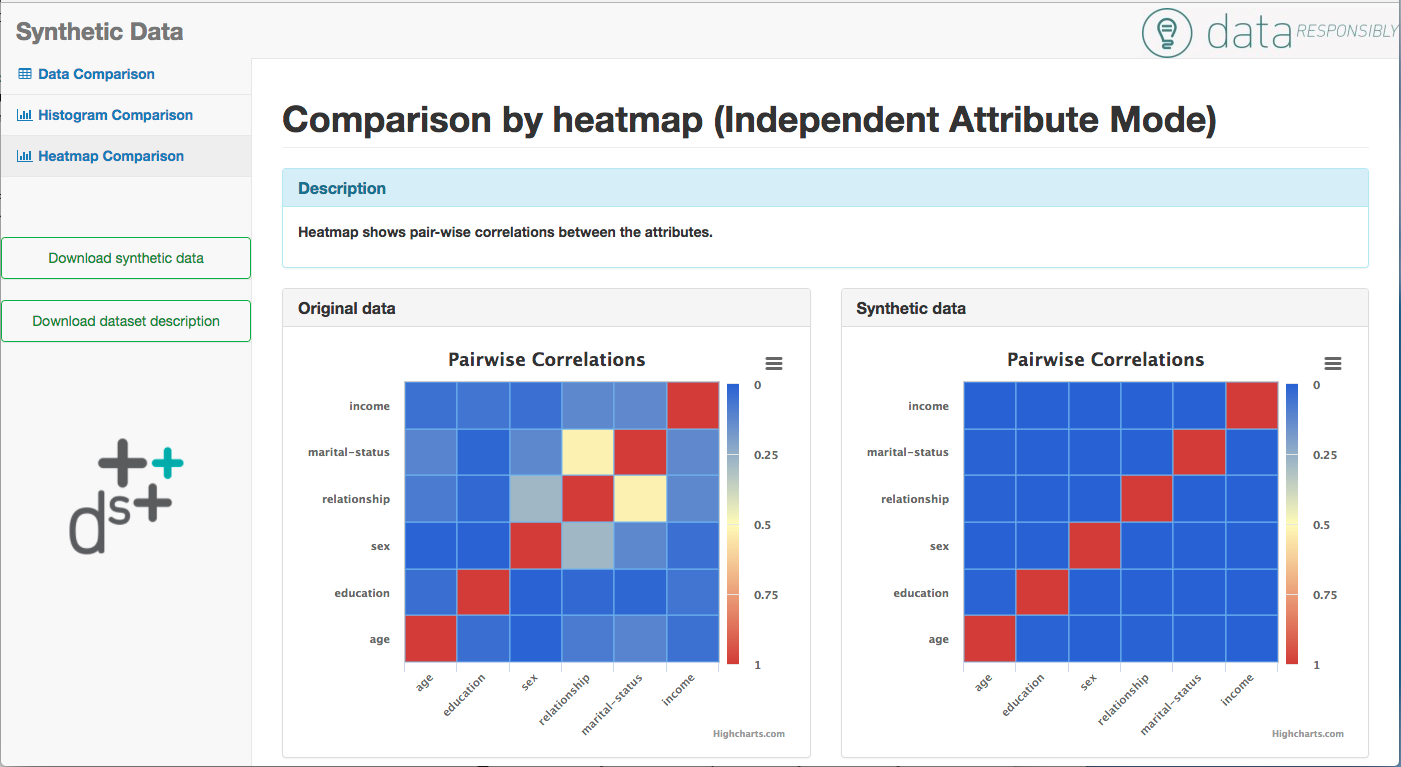}
\caption{Pair-wise correlations: independent mode.}
\label{fig:heatmap_indep}
\end{figure*}

\begin{figure*}
\centering
\includegraphics[width=5in]{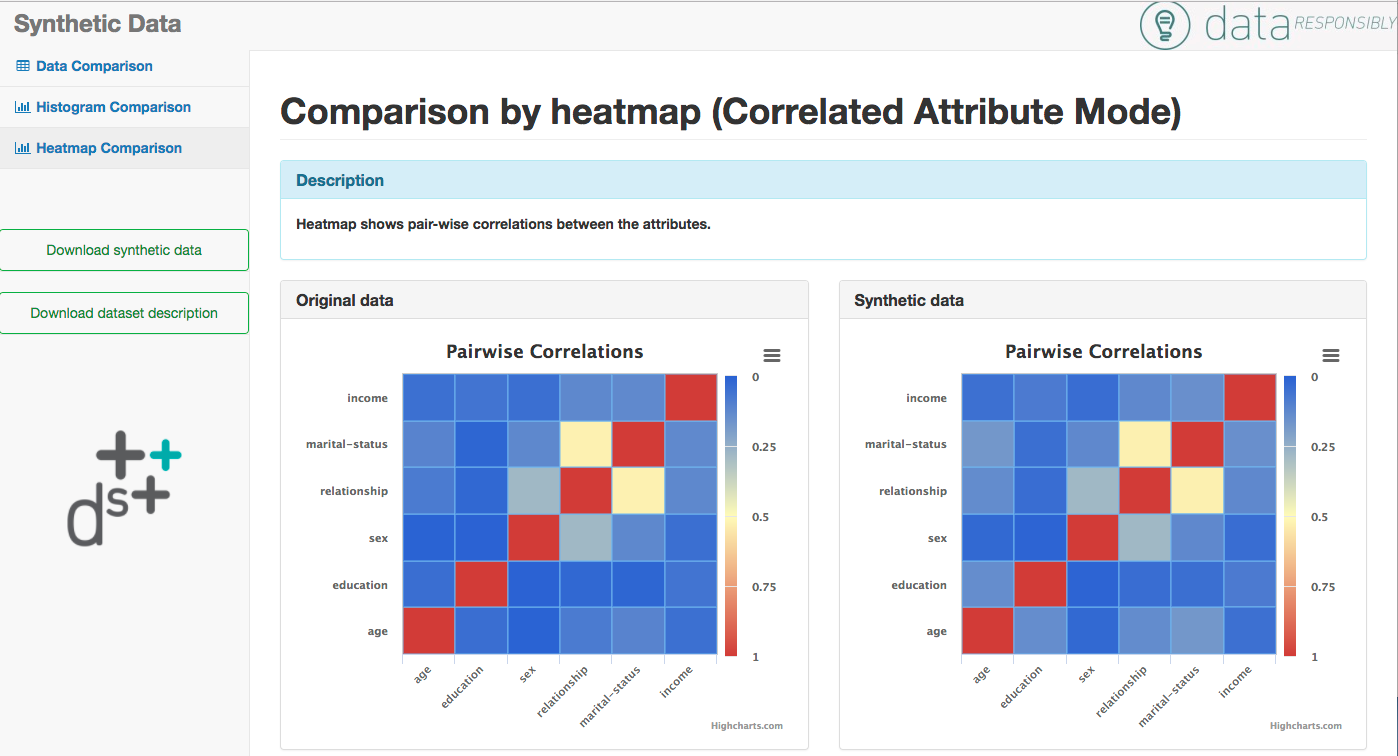}
\caption{Pair-wise correlations: correlated attribute mode.}
\label{fig:heatmap_corr}
\end{figure*}

When invoked in independent attribute mode, \ddspace implements a differentially private mechanism by adding controlled noise into the learned per-attribute distributions (histograms). The noise is from a Laplace distribution with location 0 and scale $\frac{1}{n \epsilon}$, where $n$ is the size of the input, denoted $Lap(\frac{1}{n \epsilon})$, setting $\epsilon=0.1$ by default. When Laplace noise is added to histogram frequencies, the value may become negative.  In that case the value is reset to 0~\cite{DBLP:conf/sigmod/ZhangCPSX14}. To generate a synthetic privacy-preserving dataset, the \dgspace module is invoked and generates a synthetic dataset by sampling.  Each row is sampled independently.  The value of each attribute in each row is sampled independently from the corresponding noisy histogram using uniform sampling. 

\subsubsection*{Correlated attribute mode} \label{sec:build_Bayesian_networks}

Attribute values are often correlated, e.g., \textit{age} and \textit{income} of a person. When invoked in correlated attribute mode, \ddspace uses the GreedyBayes algorithm to construct Bayesian networks (BN) to model correlated attributes~\cite{DBLP:conf/sigmod/ZhangCPSX14}.  

\begin{figure*}[t!]
\centering
\begin{minipage}{3in}
\centering
\includegraphics[width=2.8in]{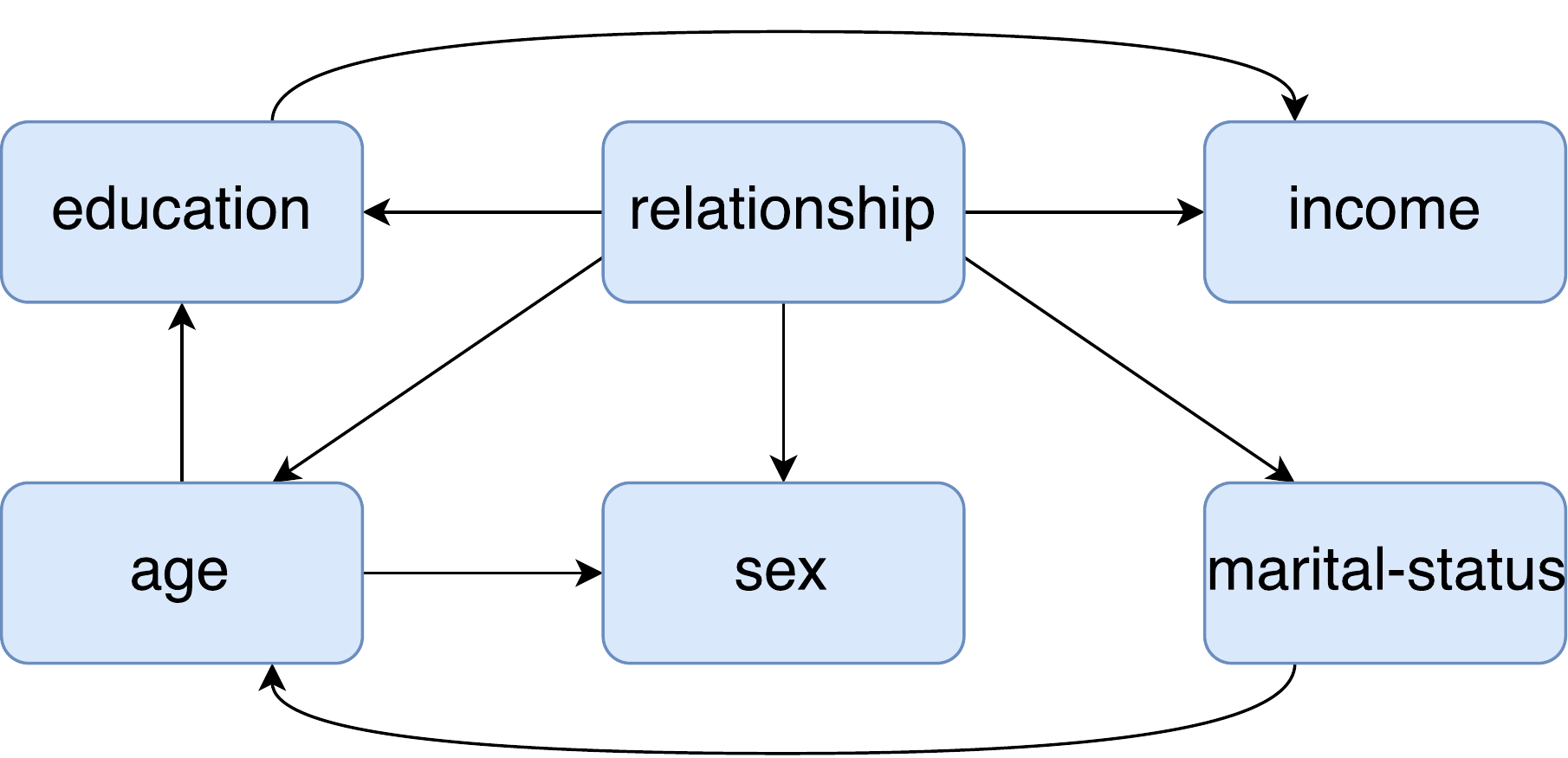}
\caption{Bayesian network: Adult Income~\cite{Lichman:2013}.}
\label{fig:bn_before}
\end{minipage}
\begin{minipage}{3in}
\centering
\includegraphics[width=2.8in]{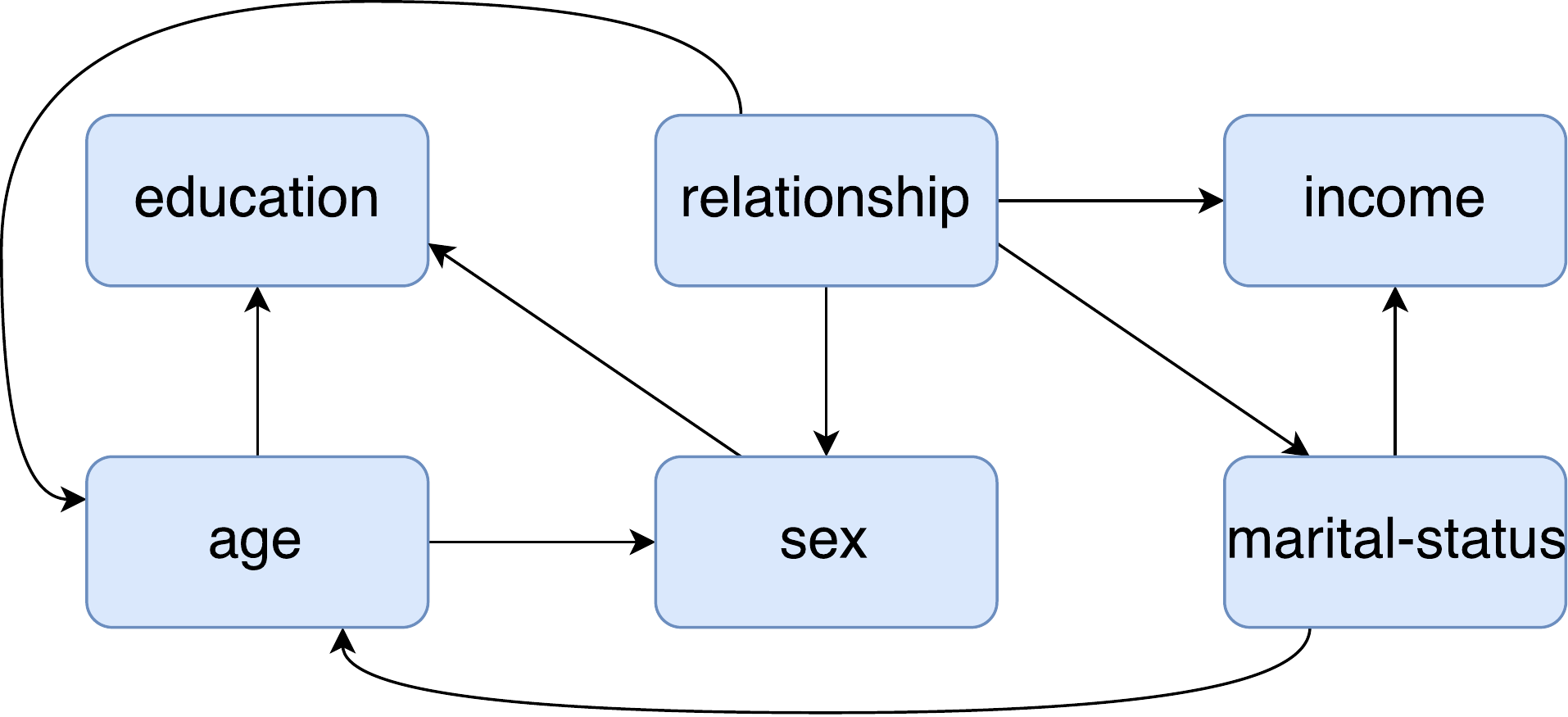}
\caption{Bayesian network: synthetic.}
\label{fig:bn_after}
\end{minipage}
\end{figure*}

The Bayesian network gives the sampling order for generating attribute values, see Figures~\ref{fig:bn_before} and~\ref{fig:bn_after} for examples.  The distribution from which a dependent attribute is sampled is called a conditioned distribution.  When constructing a noisy conditioned distribution, $Lap(\frac{4(d-k)}{n \cdot \epsilon})$ is injected to preserve privacy.  Here, $d$ is the number of attributes, $k$ is the 
maximum number of parents of a BN node, and $n$ is the number of tuples in the input dataset.  We  construct conditional distributions according to Algorithm 1 of~\cite{DBLP:conf/sigmod/ZhangCPSX14}.

The parents of a dependent attribute can be categorical or numerical, whose distributions are modeled by bar charts and histograms, respectively. The conditions for this dependent attribute are the legal values of the categorical parents and the intervals of the numerical parents. Here, the intervals are formed in the same way as the unconditioned distributions of the parent attributes. For example, the \textit{age} attribute has intervals \{[10, 20),[20, 30),[30, 40)\} in its unconditioned distribution. Assume \textit{education} only depends on \textit{age}. Its conditioned distributions will be under the same intervals, i.e., age $ \in [10,20)$, age $ \in [20,30)$ and age $ \in [30,40)$ respectively.




\subsection{Interacting with \ds}

\dsspace provides several built-in functions to inspect the similarity between the private input dataset and the output synthetic dataset. 

With the {\em Data Comparison} view (Figure~\ref{fig:datacomp}), the data owner can quickly test whether the tuples in the synthetic dataset are detectable by inspecting and comparing the raw data.  

With the {\em Comparison by histogram} view (Figure~\ref{fig:histcomp}), the user can compare the estimates of the per-attribute probability distributions in the input dataset to those in the synthetic dataset, with the expectation that these histograms will be similar in independent attribute and correlated attribute modes (as shown in Figure~\ref{fig:histcomp}), but that they would be dis-similar in random attribute mode.

With the {\em Comparison by heatmap} view (Figures~\ref{fig:heatmap_indep} and~\ref{fig:heatmap_corr}), the user can inspect pair-wise attribute correlations 
in the original dataset, and compare these to the correlations in the synthetic dataset. 
We quantify correlations using mutual information (MI), a common measure of mutual dependence between two random variables.  

Consider, for example, the comparison by heatmap in independent attribute mode presented in Figure~\ref{fig:heatmap_indep}.   Blue grid cells correspond to little or no correlation (MI close to 0), yellow cells correspond to moderate correlation (MI around 0.5), while red cells correspond to strong correlation (MI close to 1). Observe that marital status and relationship exhibit moderate correlation in the original dataset but that they do not correlate in the synthetic dataset (the corresponding cells are yellow on the left side of Figure~\ref{fig:heatmap_indep} and blue on the right side of this figure).  This effect is expected, since independent attribute mode removes any correlations between attributes.

Next, consider Figure~\ref{fig:heatmap_corr}, which presents heatmaps for correlated attribute mode.  Observe that marital status and relationship exhibit a similar level of correlation in the synthetic dataset as they did in the original.


\section{Extensions and Discussion}
\label{sec:vision}


We see synthetic data as a fundamental component of ``people analytics,'' where sensitive, private data must be used to make high-risk decisions.  Beyond the capabilities of the current \dsspace tool, we envision a number of usage scenarios and corresponding extensions; we describe these extensions in this section.  

\subsection{Enabling collaboration}

As described in the introduction, our primary motivating use case is to reduce friction in the early stages of collaboration between data providers and outside data scientists.  Our hypothesis is if data scientists are allowed to ``get their hands dirty'' with synthetic data, they are more likely to internalize the problem being solved and develop effective solutions more efficiently.  In our own experience, we find that ``whiteboard'' solutions designed prior to seeing the data often become irrelevant once the data is available --- attributes are different than we expected, data sizes are too small to train advanced models, biases in the data prevent certain kinds of analyses from taking place, inconsistent values complicate debugging (e.g., the string ``N/A'' in a column of integers).  Exposure to these challenges early helps shape the conversation and reduce effort as data sharing agreements are being prepared.  
 
\subsection{Fake Linked Data}

The value of the municipal datasets that motivate our approach enjoys a network effect: each pair of datasets enables new insights. For example, in the City of Seattle, a study is underway to determine the effect of housing instability on high school graduation rates: Do children who endure periods of homelessness graduate on time?  Although the question is simple, it involves linking two extremely sensitive datasets: student data (protected by FERPA and requiring consent of parents to use) and homelessness data (protected by HIPAA and local privacy norms).  In fact, the typical agreements governing the use of each of these datasets explicitly forbid linking them with any other datasets, and these typical agreements must be revised on a case by case basis to enable such studies.

To bootstrap collaborations over linked data, we might like to use the same approach we have described: generate fake education data, and then generate fake homelessness data.  But this na\"{i}ve will not work: synthetic records will not necessarily link with other synthetic records in a statistically similar way as in the real data.  An apparent solution is to simply join the two datasets, then generate a synthetic dataset from the result.  But this approach entails one data provider sharing their data with another provider, which is explicitly what we need to avoid.

To solve this problem, we need to estimate the statistical properties of the joined dataset, and use that information to guide the data synthesis process independently for the education and homelessness data.
We observe that there are three classes of joined tuples: housing records $H$ that have no corresponding education records, education records $E$ that have no corresponding housing record, and linked pairs of housing and education records $HE$.  We want to estimate the number in each category,  $|H|$,  $|E|$,  $|HE|$. 

To produce these estimates without sharing information, we can use locality sensitive hashing techniques~\cite{LSH} to independently map education tuples and housing tuples into a common space.  Locality sensitive hashing algorithms have the property that similar inputs are mapped to similar outputs, without coordination. For example, integers can be mapped to their most significant $n$ bits.  For structured data, one simple approach is to concatenate the values in the tuples, split this long string into n-grams, sort the n-grams lexicographically, then truncate the sorted list to retain the first $k$ n-grams.  This way, similar tuples will map to similar sequences of n-grams.

Using this approach (or more sophisticated approaches that make use of domain knowledge), we can independently map tuples from different providers into a shared space to determine how likely they are to match, and therefore estimate the counts $|H|$,  $|E|$,  $|HE|$.  Recall that $|H|$ is the number of housing tuples for which no nearby education tuples exist, and $|E|$ is the number of education tuples for which no nearby housing tuples exist.  We can assume the remaining tuples join to produce linked pairs.
Armed with these estimates, we can generate ids for education and housing tuples to ensure that an appropriate number of joined tuples are produced.  Further, we can generate attribute values guided by the same LSH techniques to ensure that joined tuples share similar values. 


\subsection{Mixing Real and Synthetic Data}

In many data sharing situations, data must be aggregated as an attempt at anonymization. Although aggregation approaches typically offer limited formal protection in practical cases~\cite{demontjoye:13,DBLP:conf/pods/DinurN03}, they are often written into data sharing policies that must be obeyed.

For example, energy providers are strongly incentivized to deliver upgrades designed to improve efficiency.  But assessing the efficacy of these upgrades using consumption data, normalized by weather and project specifications, is difficult.  Once again, the problem is to share sensitive data: the energy consumption of customers is a signal-rich resource that could be used for ``cybercasing,'' for example to predict when people leave their homes.  To mitigate such risks, the rules that govern data sharing are designed to prevent disambiguation of aggregate measures.  For instance, a geographic estimate of energy usage must aggregate no fewer than 100 consumers, and no one consumer can represent more than 10\% of the total usage.  The final calculation must require a certain number of days of recorded usage data. 

We see a novel use case for synthetic data to ``fill out'' the aggregates to meet anonymity requirements, as a kind of tuple-level imputation process.

Synthetic data may also be mixed with real data to repair global statistical anomalies, a kind of tuple-level imputation.  For example, as reported in a recent New York Times article on urban homelessness~\cite{nytimeshomeless}:
``Last year, the total number of sheltered and unsheltered homeless people in the city was 75,323, which included 1,706 people between ages 18 and 24. The actual number of young people is significantly higher, according to the service providers, who said the census mostly captured young people who received social services.'' 
This representativeness gap can be filled with synthetic data to help data scientists triage their methods, or obscure the fact that the gap exists, in case data collection activities are sensitive.

\subsection{Adversarial Fake Data Generation}

\begin{figure}[t!]
\centering
\includegraphics[width=2in]{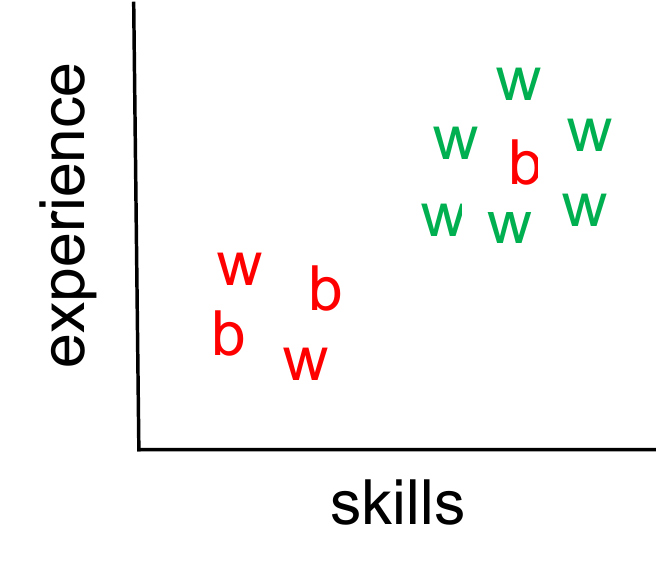}
\caption{
An illustration of a pathological dataset to evaluate disparate treatment.
The symbol \textsf{w} represents a white candidate and the symbol \textsf{b} represents a black candidate.  Red indicates that a hypothetical model rejected the candidate and green indicates that a hypothetical model accepted the candidate.  At lower left, candidates are both low-skill and low-experience, and all are rejected.  At upper right, a cluster of predominately white candidates and one black candidate receives inconsistent outcomes: if the one black candidate is rejected while similar candidates are accepted, there is evidence of disparate treatment.  Generating specific situations to test a model's response is a role that synthetic data generation can play.}
\label{fig:racist}
\end{figure}

Data providers are reluctant to share data for more than just privacy reasons. As decisions are shifted from humans to algorithms, the opportunity for, and impact of, discrimination becomes more acute.  To earn the trust of data providers and demonstrate that proposed methods are robust to biased data, we envision generating intentionally biased datasets to explore ``corner cases."  
Consider for example a hiring scenario. A cluster of job applicants who are similar in terms of experience, skills, and current position should tend to all receive offers to interview. If an African-American candidate in the cluster does not receive an offer when Caucasian candidates do, there is evidence that individual fairness is violated~\cite{DBLP:conf/innovations/DworkHPRZ12}, and, specifically in this case, that there is {\em disparate treatment} --- an illegal practice of treating an entity differently based on a protected characteristic such as race or gender. This situation is illustrated in Figure \ref{fig:racist}.  Beyond disparate treatment, adversarial synthetic datasets can be generated to test more general cases of violation of individual and group fairness, under different interpretations of these measures~\cite{DBLP:conf/innovations/DworkHPRZ12,DBLP:journals/corr/Zliobaite15a}.

Testing fairness and bias properties of algorithmic decision-making systems is particularly important in cases where black-box third party tools are used, and where intervening on the inputs and analyzing the impact of the interventions on the outputs is one of only a handful of methods to infer system behavior.  In enacting such interventions, it is particularly important to generate inputs that are realistic, and that systematically explore cases that may not have been present in the actual historical data that was used to train the model.


Existing approaches for this problem rely on random sampling of input data to measure the response of black box models \cite{DBLP:conf/sp/DattaSZ16,DBLP:journals/corr/TramerAGHHHJL15}, but random sampling cannot necessarily generate the pathological datasets that may occur in practice.

Beyond statistical bias, the ability to generate pathological datasets in terms of scale, anomalous values, and unexpected correlations can aid in debugging and stress-testing of external models, in the same way that benchmark datasets can help expose problems in, say, database systems \cite{arasu:11}.

To generate these pathological datasets, we can make a three extensions to the existing \dsspace tool: First, we can allow users to edit the distribution derived from the real data to produce extreme values.  Second, to allow even more precise control, we can design preconfigured pathological distributions to simulate, for example, individual fairness situations.  That is, using the annotations on the original data to distinguish protected attributes from non-protected attributes, we can generate clusters of similar tuples intentionally.  Third, to assess systematic robustness (as opposed to statistical robustness) we can intentionally inject pathological values into attributes --- missing values, inconsistent types, and extreme values (say, an age of -2 years).


\subsection{Synthetic Cities: Comprehensive Interconnected Administrative Datasets}

We envision combining all these techniques to generate, and incrementally improve, an administrative projection of entire virtual city to support research without data sharing encumbrances.  Unlike population synthesis approaches in urban planning \cite{farooq:13} and economics \cite{axtell:16} which use agent-based models to study the emergent dynamics of an entire city from the ground up, our approach focuses on modeling \emph{only the administrative data that would result from the dynamics of the city}, which provides a more realistic way of evaluating solutions.  Since in practice researchers will typically only have access to the administrative data, models developed based on untestable assumptions about human behavior are difficult to evaluate, and interventions based on these models are difficult to trust.  But the approaches are ultimately complementary, since artificial administrative data can be used to evaluate agent-based models, and agent-based models can be used as a source of artificial administrative data when the true datasets are not available.

\section{Related Work}
\label{sec:related}

In our work on \dsspace we leverage recent advances in practical differential privacy~\cite{DBLP:conf/sigmod/HayMMCZ16} and privacy-preserving generation of synthetic datasets~\cite{DBLP:conf/icde/LuMG14,DBLP:conf/sigmod/ZhangCPSX14}.  In particular, we make use of the privacy-preserving learning of the structure and conditional probabilities of a Bayesian network in PrivBayes~\cite{DBLP:conf/sigmod/ZhangCPSX14}, and are inspired in our implementation by the work on DPBench~\cite{DBLP:conf/sigmod/HayMMCZ16}. 

Other recent approaches in privacy-preserving data generation include the work on plausible deniability in data synthesis~\cite{DBLP:journals/pvldb/BindschaedlerSG17}, on perturbed Gibbs sampling for private data release~\cite{DBLP:conf/ichi/ParkGS13,DBLP:journals/tdp/ParkG14}, and on sampling from differentially private copula functions~\cite{DBLP:journals/pvldb/LiXZJ14}.

Data sharing systems, including SQLShare \cite{jain:16} and DataHub \cite{bhardwaj:15}, aim to facilitate collaborative data analysis, but do not incorporate privacy preserving features or purport to manage sensitive data.  We see these systems efforts as a potential delivery vector for \dsspace capabilities.

\section{Take-away messages}
\label{sec:conc}

In this paper, we argued that the generation and use of synthetic data is a critical ingredient in facilitating collaborations involving sensitive data.   The cost of establishing formal data sharing agreements limits the impact of these ad hoc collaborations in government, social sciences, health, or other areas where data is heavily encumbered by privacy rules. 

A good fake dataset has two properties: it is representative of the original data, and it provides strong guarantees against privacy violations.  

We discussed several use cases for fake data generation, and presented \ds, a privacy-preserving synthetic data generator for tabular data. Given a dataset, \dsspace can derive a structurally and statistically similar dataset, at a configurable level of statistical fidelity, while ensuring strong privacy guarantees.  \dsspace is designed with usability in mind.  The system supports three intuitive modes of operation, and requires minimal input from the user.   

We see fake data generators like \dsspace used in a variety of application contexts, both as stand-alone libraries and as components of more comprehensive data sharing platforms.  As part of ongoing work, we are studying how best to deliver these features to data owners, and determining how additional requirements can be met.  

\dsspace is open source, and is available for download at \url{https://github.com/DataResponsibly}.  The work on \dsspace is part of the Data, Responsibly project, and is a component of the Fides framework~\cite{DBLP:conf/ssdbm/StoyanovichHAMS17}, which operationalizes responsibility in data sharing, integration, analysis and use.

\balance

\bibliographystyle{ACM-Reference-Format}
\bibliography{main}

\end{document}